\documentstyle[11pt,newpasp,twoside,epsf]{article}
\def\ltsima{$\; \buildrel < \over \sim \;$}
\def\simlt{\lower.5ex\hbox{\ltsima}}
\def\gtsima{$\; \buildrel > \over \sim \;$}
\def\simgt{\lower.5ex\hbox{\gtsima}}

\def\gs{{_>\atop^{\sim}}}

\def\edcomment#1{\iffalse\marginpar{\raggedright\sl#1\/}\else\relax\fi}
\reversemarginpar

\begin{document}
\title{The Intimate Link Between Accretion and BELR}
\author{Fabrizio Nicastro}
\affil{Harvard-Smithsonian Center for Astrophysics, 60 Garden Street, 
Cambridge, MA 02138, USA}
\author{A. Martocchia}
\affil{Observatoire Astronomique, 11 Rue de l'Universit\'e, 
F--67000 Strasbourg, France}
\author{G. Matt}
\affil{Dipartimento di Fisica, Universit\`a Roma Tre, 
Via della Vasca Navale 84, I-00146 Roma, Italy}

\begin{abstract}
In this paper I present evidence suggesting that the absence or 
presence of Hidden Broad Line Regions (HBLRs) in Seyfert 2 galaxies is 
regulated by the rate at which matter accretes onto a central supermassive 
black hole, in units of Eddington. 
I discuss the above findings in the context of a model proposed 
by Nicastro (2000), in which the existence of the BLRs is regulated 
by the disk accretion rate. For low enough accretion rates, the disk 
is stable, and the BLRs do not form. 
\end{abstract}

\section{Introduction}
In the framework of orientation-based unificaton schemes (Antonucci, 1993), 
type-2 AGN are believed to be intrinsically the same as type-1 AGN, but 
seen at different angles (i.e. edge-on {\em versus} pole-on). 
This scenario was first proposed by Antonucci \& Miller 
(1985) to explain the presence of polarized broad lines in the archetypical  
Seyfert 2 (Sy2), NGC 1068, and is now supported other than by 
spectropolarimetric observations of Hidden Broad Emission Line Regions 
(HBLRs) in several other sources, also by X--ray observations, which 
demonstrate that Sy2s have usually absorption columns largely exceeding 
the Galactic ones. 

Despite observations do generally support orientation-based 
unification models for AGNs, important exceptions do exist. 
Purely orientation based unification schemes, for example, fail 
to account for all of the observed differences between two sub-classes 
of type 1 AGN: normal Seyfert 1s (Sy1s), and the so called Narrow Line 
Seyfert 1s (NLSy1s). NLSy1s have [OIII]/H$\beta$ ratio typical of type-1 
sources, but also BEL FWHM narrower than 2000 km s$^{-1}$. 
They also differ significantly from Sy1s in the X-rays, 
where they show much steeper soft (0.1-2 keV: Boller 
et al., 1996) and hard (2-10 keV: Brandt et al. 1997; Leighly, 1999a) 
continua, as well as faster variability (Leighly, 1999b). 
It has been suggested that the black hole mass, or the accretion rate, or 
both (e.g. Pounds et al. 1995; Laor et al. 1997; Nicastro, 2000: hereinafter 
N00) are responsible for the observed differences between Sy1s and NLSy1s. 

\noindent
More importantly, orientation-based models fail in their fundamental 
prediction: only about 50 \% of the brightest 
Seyfert 2s show the presence of HBLRs in their polarized optical spectra, 
while the remaining half do not (Tran, 2001; Gu \& Huang, 2002). 

Here I present evidence that suggests that the absence or 
presence of HBLRs is regulated by the ratio between the X-ray luminosity 
and the Eddington luminosity which, in the accretion--powered scenario, is a 
measure of the rate at which matter accretes onto the central supermassive 
black hole. 
This evidence is consistent with the model proposed by N00, in which the 
BLRs are formed due to accretion disk 
instabilities occurring in proximity of the critical radius at which 
the disk changes from gas pressure dominated to radiation pressure 
dominated. This radius diminishes with decreasing $\dot{m}$; for low 
enough accretion rates (and therefore luminosities), the critical radius 
becomes smaller than the innermost stable orbit, the disk is not 
longer unstable, and so BLRs do not form. Under the Keplerian assumption, 
this model is able to reproduce the whole range of observed BEL FWHM in 
AGN, as a function of the accretion rate in units of Eddington, and so 
also to explain the observed difference in BEL-FWHM between Sy1s and 
NLSy1s. 

\subsection{The Model}
The model is based on two basic assumptions: (a) a vertical disk wind, 
originating at a critical distance in the accretion disk, is the origin of 
the BEL-Clouds (BELCs), and (b) the widths of the BELs are the Keplerian 
velocities of the accretion disk at the radius where the wind arises.
The wind forms because of the existence of Lightman-Eardley 
(Lightman \& Eardley, 1974) instabilities in the radiation pressure 
dominated region of a Shakura-Sunyaev (SS) disk (Shakura \& Sunyaev, 1973), 
and so for external accretion rates higher than a minimum value below which 
a standard SS-disk is stable and extends down to the last stable orbit. 
The region of the disk where the wind forms is bound internally by the 
transition radius between the radiation pressure and gas pressure dominated 
parts of the disk, 
\begin{equation} \label{rtran}
r_{tran} f^{-16/21} \simeq 15.2 (\alpha m)^{2/21} \left({1 \over \eta} 
\dot{m} \right)^{16/21}, 
\end{equation}
and externally by the radius at which, in the context of the dynamical 
disk-corona model proposed by Witt, Czerny and Zycki in 1997 (WCZ97), 
the fraction of energy dissipated in the corona ($(1-\beta) \simeq 0.034 
(\alpha f {1 \over \eta} \dot{m})^{-1/4} r^{3/8}$) is maximum (see N00 for 
additional details)
\footnote{In the WCZ97 model a transonic vertical outflow from the disk 
develops where the fraction of total energy dissipated in the corona is 
close to unity}
:
\begin{equation} \label{rmax}
r_{max} f^{-2/3} \simeq 8,000 \left( \alpha {1 \over \eta} 
\dot{m} \right)^{2/3}. 
\end{equation}
In the above equations all quantities are dimensionless: $m = 
M/M_{\odot}$, $\dot{m} = \dot{M} / \dot{M}_{EDD}$, $r = R/R_0$, 
with $\dot{M}_{EDD} = 1.5 \times 10^{17} \eta^{-1} m \hbox{ }$ g 
s$^{-1}$, and $R_0 = 6GM/c^2$, for a non-rotating black hole; here 
$\eta$ is the maximum efficiency. Finally, $f$ gives the boundary 
conditions at the marginally stable orbit: $f = f(r) = (1 - r^{-0.5})$. 

The critical radii $r_{tran}$ and $r_{max}$ (equations 1 and 2) 
depend both on the accretion rate and, interestingly, with 
similar powers. This results in a quasi-rigid radial shifting of the region 
delimited by these two distances as the accretion rate (in critical units) 
varies. Finally, for the transonic disk outflow (and so for the BELRs), the 
model adopts an intermediated radius $r_{wind}$ computed weighting the radial 
distance between $r_{tran}$ and $r_{max}$ by $(1-\beta)$. 

Equation 1 allows us to define the minimum 
external accretion rate needed for a thermally unstable radiation 
pressure dominated region to exist. 
From the condition $r > 1.36$ (the limit of validity of the SS-disk 
solution) we have: $\dot{m} \gs \dot{m}_{min}(m) 
\simeq 0.3 \eta (\alpha m)^{-1/8}$. 
Throughout this paper I assume $\eta = 0.06$, and a viscosity 
coefficient of $\alpha = 0.1$, which give a minimum external accretion 
rate of $\dot{m}_{min} \sim (1-4) \times 10^{-3}$, for $m$ in 
the range $10^6 - 10^9$ (see Fig. 1 of N00). 
At lower accretion rates a SS-disk (SS73) is stable down to the 
last stable orbit: we propose that all the available 
energy is dissipated in the disk and no radiation pressure 
supported and driven wind is generated. AGN accreting at these low 
external rates should show no BELs in their optical spectra. 

\section{Testing the Model at Low Accretion Rates}
One of the consequences of the N00 model predictions is 
that a fraction of optically classified Sy2s should exist, which show 
no BELs in polarized light: all those source accreting at rates lower 
than the minimum allowed rate for the formation of BELRs. 
To test this prediction, we extracted a sub-sample of Sy2s from the 
Tran (2001; 'primary' sample) and Gu \& Huang (2002; 'secondary' sample) 
spectropolarimetric samples, which were observed in the X-rays at least 
once with imaging X-ray satellites (Nicastro, Martocchia \& Matt, 2003: 
NMM03). The goal was to derive reliable X-ray (i.e. 'nuclear') luminosities 
and central black hole masses for the sources of our sample, to get an 
estimate of the external accretion rate in units of Eddington. 
Details on the sample selection criteria and the techniques used to 
derive the black hole masses can be found in NMM03. 
Our final sample containes 15 Sy2s, 9 HBLRs and 6 non-HBLRs. 

Results are summarized in Figure 1 that shows fractional luminosities 
in units of Eddington, versus black hole masses for all the 15 sources of 
our sample. We first note that a very broad range of accretion rates is 
spanned by the sources of our sample (more than three orders of magnitude), 
which are otherwise powered by central black holes with rather homogeneous 
masses (only a factor of about 15 across the entire sample). 
Most importantly, Figure 1 clearly shows that HBLR sources are accreting 
at much faster rates compared to non-HBLR sources. 

\noindent
The threshold value of $\dot{m}_{thres} \simeq 10^{-3}$ divides up HBLR from 
non-HBLR sources in the $M_{BH}$ vs $\dot{m}$ plane (dashed vertical line 
in Fig. 1). 
The only exceptions are NGC~3081 and NGC~3281, both sources from our 
``secondary'' sample (but see N00 for details and comments on the 
reliability of the X-ray and spectropolarimetric optical data of these 
two sources). 
%
%%%%%%%%%%%%%%
\begin{figure}
\epsfysize=2.5in 
\epsfxsize=2.5in 
\hspace{1.5in}\epsfbox{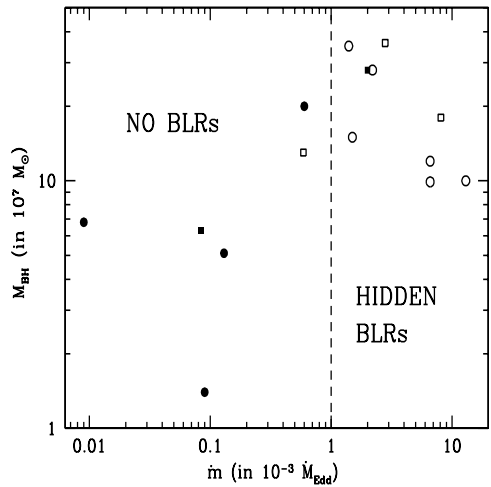} 
%\plotfiddle{f1.eps}{2in}{0}{}{}{}{}
\caption{Black hole masses vs accretion rates (defined as $\dot{m} = 
\dot{M}/\dot{M_{Edd}} = L_{bol}/L_{Edd}$). Open and filled symbols 
refer to HBLRs and non-HBLRs. Circles and squares are sources from our 
primary and secondary samples, respectively (see N00 for details).}
\end{figure}
%%%%%%%%%%%%%%
%

\section{Conclusion}
In this contribution I presented a model that links the physical 
mechanism responsible for the AGN activity (i.e. the accretion) with 
the existence and width of the BELs. This model seems to naturally 
(a) reproduce the Sy1-{\em versus}-NLSs1 observed differences, and (b) 
account for those 50 \% of Sy2s for which no HBLRs has been observed.  
Additional testing of this model is needed, and requires 
larger samples both in the optical (e.g. SDSS photometry and spectra), 
and in the X-rays (e.g. {\em Chandra} and {\em Newton}-XMM spectra).

\end{document}